\documentclass[conference]{IEEEtran}
\usepackage[latin9]{inputenc}
\usepackage{amsmath}
\usepackage{amssymb}
\usepackage{graphicx}
\usepackage{esint}

\makeatletter
\usepackage{array}
\usepackage{url}
\usepackage{multirow}
\usepackage{cuted}

\DeclareTextSymbolDefault{\textquotedbl}{T1}
\IEEEoverridecommandlockouts
\usepackage{cite}
\usepackage{amsfonts}
\usepackage{algorithmic}
\usepackage{textcomp}
\usepackage{xcolor}
\def\BibTeX{{\rm B\kern-.05em{\sc i\kern-.025em b}\kern-.08em
		T\kern-.1667em\lower.7ex\hbox{E}\kern-.125emX}}

\makeatother

\begin{document}
\title{An EM Body Model for Device-Free Localization with Multiple Antenna Receivers: A First Study
\thanks{Funded by the European Union. Views and opinions expressed are however those of the author(s) only and do not necessarily reflect those of the European Union or European Innovation Council and SMEs Executive Agency (EISMEA). Neither the European Union nor the granting authority can be held responsible for them. Grant Agreement No: 101099491.}}
\author{\IEEEauthorblockN{Vittorio Rampa\textit{$^{1}$}, Federica Fieramosca\textit{$^{2}$},
Stefano Savazzi\textit{$^{1}$}, Michele D'Amico\textit{$^{2}$}} \IEEEauthorblockA{\textit{$^{1}$}\textit{\emph{ Consiglio Nazionale delle Ricerche}}\emph{,}
\textit{\emph{IEIIT}} institute, Piazza Leonardo da Vinci 32, I-20133,
Milano, Italy.\\
 \textit{$^{2}$}\textit{\emph{ }}DEIB, Politecnico di Milano, Piazza Leonardo
da Vinci 32, I-20133, Milano, Italy\linebreak{}
 } }
\maketitle
\begin{abstract}
Device-Free Localization (DFL) employs passive radio techniques capable
to detect and locate people without imposing them to wear any electronic
device. By exploiting the \emph{Integrated Sensing and Communication}
paradigm, DFL networks employ Radio Frequency (RF) nodes to measure
the excess attenuation introduced by the subjects (i.e., human bodies) moving
inside the monitored area, and to estimate their positions and movements.
Physical, statistical, and ElectroMagnetic (EM) models have been proposed
in the literature to estimate the body positions according to the
RF signals collected by the nodes. These body models usually employ
a single-antenna processing for localization purposes. However, the
availability of low-cost multi-antenna devices such as those used
for WLAN (Wireless Local Area Network) applications and the timely
development of array-based body models, allow us to employ array-based
processing techniques in DFL networks. By exploiting a suitable array-capable
EM body model, this paper proposes an array-based framework
to improve people sensing and localization. In particular, some simulations
are proposed and discussed to compare the model results in
both single- and multi-antenna scenarios. The proposed framework paves
the way for a wider use of multi-antenna devices (e.g., those employed
in current IEEE 802.11ac/ax/be and forthcoming IEEE 802.11be networks)
and novel beamforming algorithms for DFL scenarios. 
\end{abstract}


\section{Introduction}

\label{sec:intro} Device-Free Localization (DFL), also known as Passive
Radio Localization (PRL), is an opportunistic set of methods capable
to detect, locate, and track people in a monitored area covered by
ambient Radio Frequency (RF) signals. By exploiting the \emph{Integrated
Sensing and Communication} paradigm~\cite{savazzi-2016}, DFL systems
are able to transform each RF node of the wireless network covering
the monitored area into a \emph{virtual sensor} able perform sensing
operations as well.

For instance, the RF signals emitted for communication purposes by
a wireless network over the monitored area, can be usefully employed
to estimate people location and status information from the received
ElectroMagnetic (EM) field. In fact, it is well known that the presence,
and the movements, of people, or objects, generically indicated as \emph{targets},
induce modifications and alterations of the EM field~\cite{koutatis-2010}
collected by each wireless device of the communication network. These
perturbations not only impair the radio channel, but can be also measured
and processed to estimate information about the presence, location,
and status (e.g., posture, movements, size, etc.) of the targets~\cite{shit-2019}.

Body models for DFL applications have been mostly evaluated for single-antenna
DFL systems for both single-target~\cite{wilson-2010,eleryan-2011,mohamed-2017,rampa-2017}
and multi-targets~\cite{rampa-2022} scenarios. These models have
been applied to different radio channel measurements, such as Channel
State Information (CSI), Received Signal Strength (RSS), Angles of
Arrival (AoA) and Time of Flight (ToF)~\cite{savazzi-2016}, and
processed by different methods such as radio imaging~\cite{wilson-2010},
Bayesian tracking~\cite{rampa-2021}, fingerprinting methods~\cite{savazzi-2016},
Compressive Sensing algorithms~\cite{wang-2012}, and Machine Learning/Deep
Learning (ML/DL) systems~\cite{shit-2019,sukor-2020}, to evaluate
the connection between targets location and the corresponding
measured perturbations.

At the same time, the wide diffusion of multi-antenna WLAN devices
(e.g., WiFi devices built according to the IEEE 802.11n/ac/ax standards
and the forthcoming IEEE 802.11be one), and the concomitant availability
of CSI extraction tools \cite{halperin-2011,xie-2018,atif-2020},
have stimulated the research activities with the adoption of multi-antenna
CSI-based processing systems for DFL applications \cite{zhang-2018,shukri-2019,garcia-2020}.

Unfortunately, only a few body models deal with antenna arrays~\cite{ruan-2016,rampa-2022b,ojeda-2022}
since almost all array-based DFL systems hold the attention on the
processing steps instead of focusing on the body model. However, reference~\cite{ruan-2016}
employs a very simple propagation model while~\cite{ojeda-2022}
exploiting a computational-intensive Ray Tracing (RT) approach. Of course,
EM simulators can be also used, but they are too complex to be of practical
use for on-line DFL systems as based on RT techniques, uniform theory
of diffraction (UTD), or full-wave methods as shown in~\cite{rampa-2022}
(and included references).

The paper extends the physical-statistical array-based model proposed
by the authors in~\cite{rampa-2022b} to predict the body-induced
propagation losses using WLAN nodes equipped with multiple antennas
at the receiver node. The body model is then used, in conjunction with
array processing techniques, to focus on the AoA and the position of
the target. The paper is organized as follows: in Sect.~\ref{sec:Array-based-body-model},
the array-based electromagnetic body model for applications in multi-antennas
DFL systems is briefly recalled and discussed. Sect.~\ref{sec:array-processing}
proposes the array-based processing tools for DFL applications while
Sect.~\ref{sec:preliminary_results} presents some preliminary simulation results comparing the array-based model with the single-antenna one.
Finally, in Sect.~\ref{sec:conclusions}, we draw some preliminary conclusions
and propose future activities. 
\noindent \begin{center}
\begin{figure}[t]
\begin{centering}
\includegraphics[scale=0.35]{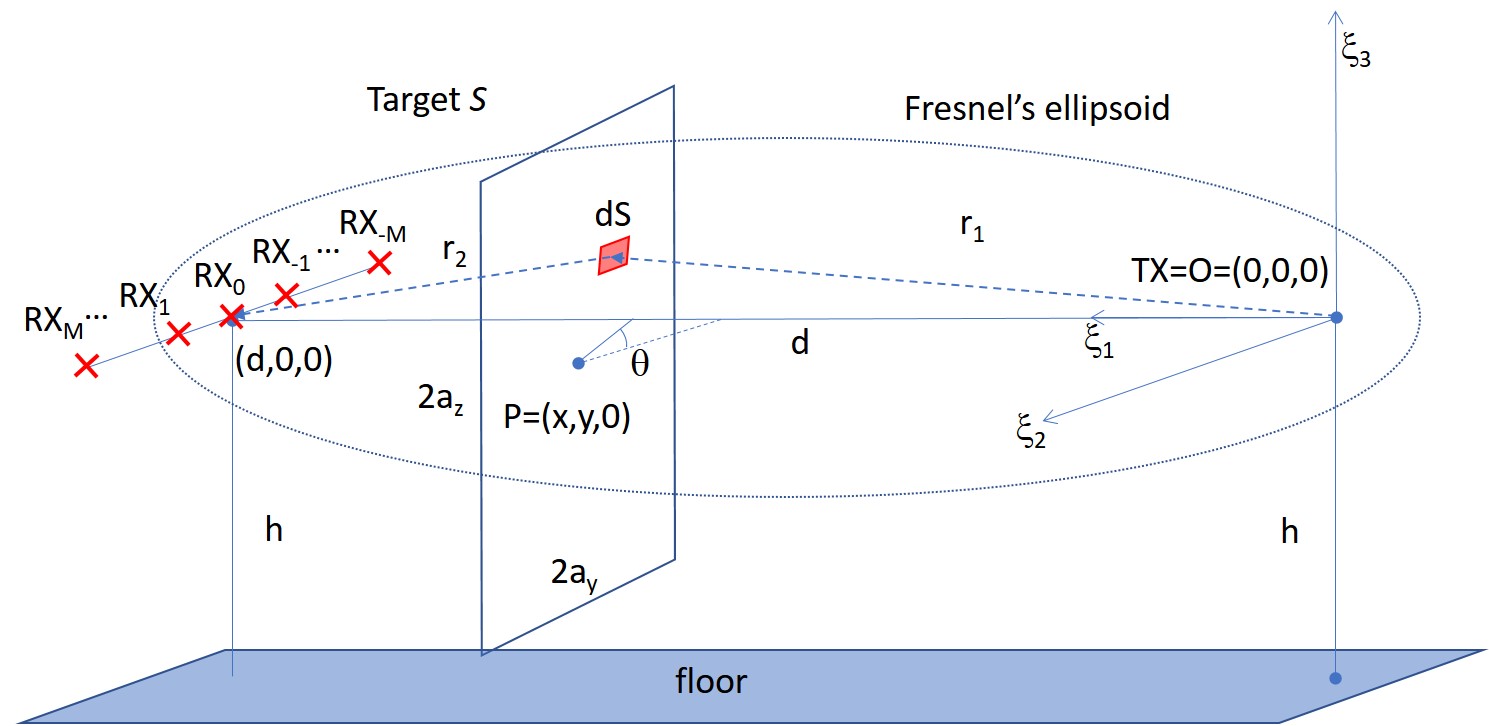} 
\par\end{centering}
\caption{3-D deployment of the radio link using an ULA-like array of $2M+1$
antennas. The segment where the array is deployed is placed at distance
$d=d_{0}$ from the TX and it is orthogonal to the LoS path connecting
the $TX$ node with the $RX_{0}$ one.}
\label{fig:array_layout} 
\end{figure}
\par\end{center}

\section{EM body model with multiple antenna receiver}
\label{sec:Array-based-body-model}

In this section we briefly recall the body model proposed in~\cite{rampa-2022b}
for a single link scenario where an antenna array is used at the receiver
while only one transmitter (TX) is used. Fig.~\ref{fig:array_layout}
shows the layout for an Uniform Linear Array (ULA) of $2M+1$ isotropic
receiver antennas RX$_{m}$, with$-M\leq m\leq M$ and $M$ an integer,
when a single target $S$ is in the monitored area near the radio
link.

The 2-D footprint of the 3-D deployment of Fig.~\ref{fig:array_layout}
is also shown in Fig.~\ref{fig:array_layout-1}, where each \emph{m}-th
antenna RX$_{m}$ of the array is uniformly deployed at mutual distance
$d_{a}$ along a segment orthogonal to the Line-of-Sight (LoS) at
distance $d=d_{0}$ from the TX and horizontally placed w.r.t. the
floor. The central antenna is indicated by the index $m=0$.

Without any loss of generality, we assume here that the floor has
no influence on the radio propagation but it only needed to support
the standing target. However, if needed, the EM influence of the floor
can be easily included as shown in~\cite{fieramosca-2023}. The target
is sketched~\cite{rampa-2017} as a vertical standing absorbing 2-D sheet $S$
of height $2a_{z}$ and traversal size $2a_{y}$ that is rotated of
the angle $\theta$ w.r.t. the $\xi_{2}$ axis. Excluding mutual antenna
coupling (approximately valid for $d_{a}>\lambda/4$), the electric
field $E^{\left(m\right)}$ received by the \emph{m}-th antenna of
the array, is:

\begin{equation}
\begin{array}{ll}
\frac{E^{\left(m\right)}}{E_{R}^{\left(m\right)}}= & 1-j\frac{d_{m}}{\lambda}\intop_{S}\frac{1}{r_{1,m}r_{2,m}}\cdot\\
 & \cdot\exp\left\{ -j\frac{2\pi}{\lambda}\bigl(r_{1,m}+r_{2,m}-d_{m}\bigr)\right\} d\xi_{2}\,d\xi_{3},
\end{array}\label{eq:E_full_array}
\end{equation}
where $E_{R}^{\left(m\right)}$ is the EM field received by the same
$RX_{m}$ node in the reference condition i.e. the free-space scenario
characterized by the absence of any target in the link area. The term
$d_{m}$ indicates the distance of the \emph{m}-th antenna $RX_{m}$
of the array from the TX while $d_{1,m}$ and $d_{2,m}$ are the distances
of the projection point $O_{m}^{'}$ (of the barycenter $P$ of the
2-D surface $S$) from the TX and $RX_{m}$ nodes. Likewise, $r_{1,m}$
and $r_{2,m}$ are the distances of the generic elementary area $dS$
of the target $S$ from the TX and $RX_{m}$, respectively. In (\ref{eq:E_full_array}),
the integration operations are performed over the squared domain $S$
having height $2a_{z}$ and traversal size $2a_{y}$.

Notice that, for $M=0$, equation (\ref{eq:E_full_array}) reduces
to the single-antenna case where RX$_{0}$ coincides with the RX antenna
at distance $d=d_{0}$ from the TX. The excess attenuation~\cite{rampa-2017}
measured by the m-th antenna of the array, that is due to the target
w.r.t. the free-space scenario, is thus given by $A_{T}^{\left(m\right)}=P_{R}^{\left(m\right)}/P^{\left(m\right)}=\left|E_{R}^{\left(m\right)}/E^{\left(m\right)}\right|^{2}$
where $P_{R}^{\left(m\right)}$ and $P^{\left(m\right)}$ are the
\emph{m}-th received power in without and with a target in the link
area, respectively. Usually, the excess attenuation is given in dB
as $A_{T,dB}^{\left(m\right)}=10\,\log_{10}\left|E_{R}^{\left(m\right)}/E^{\left(m\right)}\right|^{2}$.
It is worth noticing that, for short arrays, the excess attenuation
due to the target is almost constant along the receiving antennas
as shown in~\cite{rampa-2022b}.

In free-space conditions, the term $E_{R}^{\left(m\right)}/E_{R}^{\left(0\right)}$
is given by:

\begin{equation}
\begin{array}{ll}
\frac{E_{R}^{\left(m\right)}}{E_{R}^{\left(0\right)}}= & \frac{d_{0}}{d_{m}}\,\exp\left\{ -j\frac{2\pi}{\lambda}\bigl(d_{m}-d_{0}\bigr)\right\} ,\end{array}\label{eq:E_R_array_0}
\end{equation}
where $E_{R}^{\left(0\right)}$ is the electric field received by
the central antenna of the array of index $m=0$ that is on the LoS
at distance $d_{0}$ from the TX.
Thus, (\ref{eq:E_full_array}) can be rearranged as:

\begin{equation}
\begin{array}{ll}
\frac{E^{\left(m\right)}}{E_{R}^{\left(0\right)}}= & \frac{d_{0}}{d_{m}}\,\exp\left\{ -j\frac{2\pi}{\lambda}\left(d_{m}-d_{0}\right)\right\} \bigl[1-j\frac{d_{m}}{\lambda}\intop_{S}\frac{1}{r_{1,m}r_{2,m}}\cdot\bigr.\\
 & \bigl.\cdot\exp\left\{ -j\frac{2\pi}{\lambda}\left(r_{1,m}+r_{2,m}-d_{m}\right)\right\} d\xi_{2}\,d\xi_{3}\bigr].
\end{array}\label{eq:E_full_array_new}
\end{equation}
For additional details about the
array-based model (\ref{eq:E_full_array_new}) and its validation with
the commercial EM simulator FEKO~\cite{elsherbeni-2014}, the interested
reader can refer to~\cite{rampa-2022b}. 
\noindent \begin{center}
\begin{figure}[t]
\begin{centering}
\includegraphics[scale=0.5]{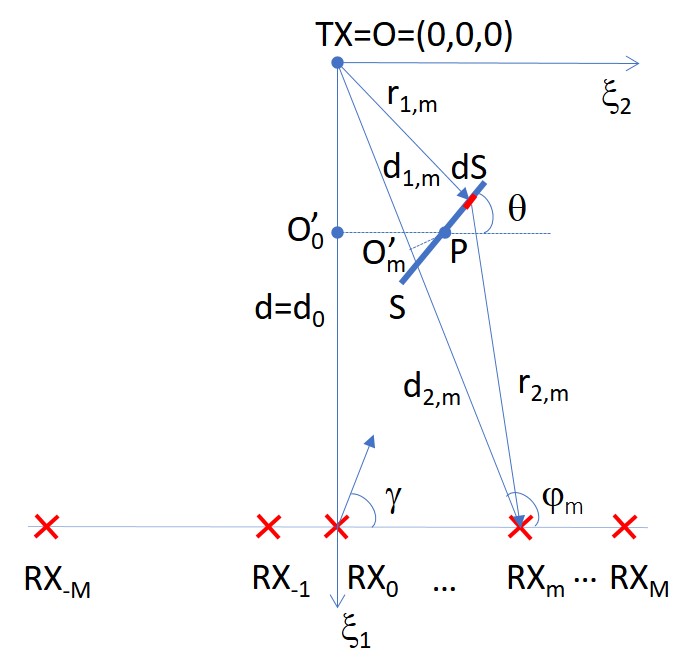} 
\par\end{centering}
\caption{\label{fig:array_layout-1}2-D layout of the radio link of Fig.~\ref{fig:array_layout}.
The point $O_{m}^{'}$ is the projection of the barycenter $P$ of
the target $S$ over the \emph{m}-th LoS path having length $d_{m}$.}
\end{figure}
\par\end{center}

\section{Array processing with model predictions}

\label{sec:array-processing}

In the following sections, we consider the ULA-arranged layout of
$2M+1$ antennas of Figs.~\ref{fig:array_layout} and~\ref{fig:array_layout-1}.
Classical array processing, such as conventional beamforming~\cite{benesty-2021}, is designed to \emph{steer}
the antenna array in one direction by forming a linear combination
of each antenna contribution. Being $\left(\cdot\right)^{T}$ the
vector transposition operator, the output of the beamforming processing,
considering $2M+1$ antennas and the vector $\mathbf{w}=\left[w_{-M}\,...\,w_{-1}\,w_{0}\,w_{1}\,...\,w_{M}\right]^{T}$
of size $2M+1$ of linear beamforming coefficients, is thus given
in general by:

\begin{equation}
y\left(t,\mathcal{S}\right)=\sum_{m=-M}^{+M}w_{m}^{*}\,r_{m}\left(t,\mathcal{S}\right)=\mathbf{w}^{H}\mathbf{r}\left(t,\mathcal{S}\right),\label{eq:beam_forming}
\end{equation}

\noindent where $\left(\cdot\right)^{*}$ indicates the conjugate operator and $\left(\cdot\right)^{H}$ the Hermitian
operator, while $r_{m}\left(t,\mathcal{S}\right)$ is the received
EM field at the $RX_{m}$ antenna. The column vector $\mathbf{r}\left(t,\mathcal{S}\right)$
of size $2M+1$, collects all the received electric field terms $\mathbf{r}\left(t,\mathcal{S}\right)=\left[r_{-M}\left(t,\mathcal{S}\right)\,...r_{-1}\left(t,\mathcal{S}\right)\,r_{0}\left(t,\mathcal{S}\right)\,r_{1}\left(t,\mathcal{S}\right)\,...\,r_{M}\left(t,\mathcal{S}\right)\right]^{T}$.
Focusing on passive sensing applications, two main scenarios are adopted
for processing: the first one refers to the empty environment (i.e.,
the free-space case, namely $\mathcal{S}=0$) while, in the second
scenario, the target $S$ is present in the monitored area ($\mathcal{S}=1$).
According to this assumption, the \emph{m}-th component of the received
vector $\mathbf{r}\left(t\right)$ is defined as:

\noindent 
\begin{equation}
r_{m}\left(t,\mathcal{S}\right)=\left\{ \begin{array}{ll}
E_{R}^{\left(m\right)}+n_{m}=E_{R}^{\left(0\right)}\,\frac{E_{R}^{\left(m\right)}}{E_{R}^{\left(0\right)}}+n_{m} & \textrm{if \ensuremath{\mathcal{S}=0}}\\
E^{\left(m\right)}+n_{m}=E_{R}^{\left(0\right)}\,\frac{E_{R}^{\left(m\right)}}{E_{R}^{\left(0\right)}}\,\frac{E^{\left(m\right)}}{E_{R}^{\left(m\right)}}+n_{m} & \textrm{if \ensuremath{\mathcal{S}=1}}
\end{array}\right.\label{eq:signal_model_reduced}
\end{equation}
where $n_{m}$ is the \emph{m}-th component of the Additive White
Gaussian Noise (AWGN) complex vector $\mathbf{n}\left(t\right)=\left[n_{-M}\left(t\right)\,...n_{-1}\left(t\right)\,n_{0}\left(t\right)\,n_{1}\left(t\right)\,...\,n_{M}\left(t\right)\right]^{T}$
of size $2M+1$, that is assumed to be spatially white with zero mean
and covariance $\sigma_{n}^{2}\mathbf{I}$, while $E_{R}^{\left(m\right)}/E_{R}^{\left(0\right)}$
and $E^{\left(m\right)}/E_{R}^{\left(0\right)}$ are given by (\ref{eq:E_R_array_0})
and (\ref{eq:E_full_array_new}), respectively. We assume here that
the noise distribution does not change according to the presence or
absence of the target. For discussions about this topic, the interested
reader can refer to~\cite{rampa-2017,rampa-2021,rampa-2022}. As
analyzed in the following sections, different beamforming methods
correspond to different choices of the weighting vector $\mathbf{w}$.

Considering the passive radio sensing problem, we are interested in
evaluating the power $P_{y}\left(\mathcal{S}\right)$ of the signal
$y$ with or without target $S$. This is given by:

\begin{equation}
\begin{array}{ll}
P_{y}\left(\mathcal{S}\right) & =\mathbb{E}\left[y\left(t,\mathcal{S}\right)\,y^{*}\left(t,\mathcal{S}\right)\right]\\
 & =\mathbf{w}^{H}\,\mathbf{R}\left(\mathcal{S}\right)\,\mathbf{w}\,
\end{array}\label{eq:beam_forming_power}
\end{equation}
with $\mathbf{R}\left(\mathcal{S}\right)=\mathbf{R}=\mathbb{E}\left[\mathbf{r}\left(t\right)\,\mathbf{r}^{H}\left(t\right)\right]$
of size $\left(2M+1\right)\times\left(2M+1\right)$, being the autocorrelation
matrix of $y\left(t,\mathcal{S}\right).$

\subsection{Impact of beamforming on body induced attenuation}

In what follows, we model the received EM field $r_{m}\left(t,\mathcal{S}\right)$
in (\ref{eq:beam_forming}) using the diffraction model considerations
and exploiting the equations (\ref{eq:E_full_array})-(\ref{eq:E_full_array_new})
for $\mathcal{S}=1$, and (\ref{eq:E_R_array_0}) for $\mathcal{S}=0$.
For conventional ULA scenarios, that assume planar wavefront propagation,
the steering (column) vector $\mathbf{a}=\left[a_{m}\right]$ of size
$2M+1$ for the considered array is given by~\cite{benesty-2021}:

\begin{equation}
\begin{array}{ll}
\mathbf{a}= & \left[\exp\left\{ -jM\frac{2\pi}{\lambda}d_{a}\cos\gamma\right\} \right.\\
 & \:\exp\left\{ -j\left(M-1\right)\frac{2\pi}{\lambda}d_{a}\cos\gamma\right\} \\
 & ...\:1\:...\\
 & \:\exp\left\{ +j\left(M-1\right)\frac{2\pi}{\lambda}d_{a}\cos\gamma\right\} \\
 & \left.\exp\left\{ +jM\frac{2\pi}{\lambda}d_{a}\cos\gamma\right\} \right]^{T},
\end{array}
\label{eq:steering_vector}
\end{equation}
where $\gamma$ is also known as DoA (Direction Of Arrival): this
is the direction of propagation of the impinging wavefront w.r.t.
the axis of the array, and $d_{a}$ is the inter-element antenna distance.
According to (\ref{eq:steering_vector}), it is also $\left|\mathbf{a}\right|^{2}=\mathbf{a}^{H}\mathbf{a}=(2M+1)$.
However, due to the fact that most DFL applications are employed in indoor scenarios with relatively short links compared with the usual far-field hypothesis, the planar wavefront assumption no longer holds and the new steering vector is:

\begin{equation}
\begin{array}{ll}
\mathbf{a}= & \left[\frac{d_{0}}{d_{-M}}\,\exp\left\{ -jM\frac{2\pi}{\lambda}d_{a}\frac{\cos\left\{ \left(\gamma+\varphi_{-M}\right)/2\right\} }{\cos\left\{ \left(\gamma-\varphi_{-M}\right)/2\right\} }\right\} \right.\\
 & \frac{d_{0}}{d_{-M+1}}\,\exp\left\{ -j\left(M-1\right)\frac{2\pi}{\lambda}d_{a}\frac{\cos\left\{ \left(\gamma+\varphi_{-M+1}\right)/2\right\} }{\cos\left\{ \left(\gamma-\varphi_{-M+1}\right)/2\right\} }\right\} \\
 & \:...\:1...\:\\
 & \frac{d_{0}}{d_{M-1}}\,\exp\left\{ +j\left(M-1\right)\frac{2\pi}{\lambda}d_{a}\frac{\cos\left\{ \left(\gamma+\varphi_{M-1}\right)/2\right\} }{\cos\left\{ \left(\gamma-\varphi_{M-1}\right)/2\right\} }\right\} \\
 & \left.\frac{d_{0}}{d_{M}}\,\exp\left\{ +jM\frac{2\pi}{\lambda}d_{a}\frac{\cos\left\{ \left(\gamma+\varphi_{M}\right)/2\right\} }{\cos\left\{ \left(\gamma-\varphi_{M}\right)/2\right\} }\right\} \right]^{T},
\end{array}\label{eq:new_steering_vector}
\end{equation}
where the angle $\varphi_{m}$, formed by the LoS of the \emph{m}-th
array antenna with the $\xi_{2}$ axis, also verifies $\forall m=0,\pm1,...,\pm M$
the following equation $\varphi_{m}=\arccos\left[\left(d_{0}/d_{m}\right)\,\cos\gamma-m\,\left(d_{a}/d_{m}\right)\right]$
with $\varphi_{0}=\gamma$. Moreover, it is also $\left|\mathbf{a}\right|^{2}=\sum_{m=-M}^{M}\left(d_{0}/d_{m}\right)^{2}$.
It is easy to check that (\ref{eq:new_steering_vector}) converges
to (\ref{eq:steering_vector}) for planar wavefronts i.e. when $\forall m$,
it is $\varphi_{m}=\gamma$. According to (\ref{eq:new_steering_vector}),
the received field $\mathbf{r}\left(t,\mathcal{S}\right)$ is given by:

\begin{equation}
\mathbf{r}\left(t,\mathcal{S}\right)=\left\{ \begin{array}{ll}
E_{R}^{\left(0\right)}\,\mathbf{a}+\mathbf{n} & \textrm{if \ensuremath{\mathcal{S}=0}}\\
E_{R}^{\left(0\right)}\,\textrm{diag\ensuremath{\left(\mathbf{a}\right)}}\,\mathbf{E}_{r}+\mathbf{n}= & \textrm{if \ensuremath{\mathcal{S}=1},}
\end{array}\right.\label{eq:signal_model}
\end{equation}
where the column vector $\mathbf{E}_{r}$ of size $2M+1$ represents the electric field ratio (\ref{eq:E_full_array}) received by the antenna array for $\ensuremath{\mathcal{S}=1}$. The $m$-th element of this vector is equal to $E^{\left(m\right)}/E_{R}^{\left(m\right)}$.

According to diffraction model of Sect.~\ref{sec:Array-based-body-model}, and by rearranging the terms in (\ref{eq:signal_model}) with (\ref{eq:new_steering_vector}) and (\ref{eq:E_full_array}), the received field $y\left(t,\mathcal{S}\right)$ after linear beamforming becomes as in eq. (\ref{eq:y_t}). Similarly as for the single antenna case~\cite{rampa-2017}, the mean excess attenuation $A_{T}$ due to the target $S$ w.r.t. the empty environment, corresponds to the ratio between $P_{y}\left(\mathcal{S}=0\right)$ and $P_{y}\left(\mathcal{S}=1\right)$, by neglecting the noise terms:

\begin{equation}
\begin{array}{ll}
A_{T}= & \frac{P_{y}\left(\mathcal{S}=0\right)}{P_{y}\left(\mathcal{S}=1\right)}=\frac{\mathbf{w}^{H}\,\mathbf{a}\,\mathbf{a}^{H}\mathbf{w}}{\mathbf{w}^{H}\,\textrm{diag}\left(\mathbf{a}\right)\,\mathbf{R}_{E}\,\textrm{diag}\left(\mathbf{a}^{H}\right)\mathbf{w}}.\end{array}\label{eq:beam_forming_attenuation}
\end{equation}
In particular, $R_{E}$ is the autocorrelation matrix of $E^{\left(m\right)}/E_{R}^{\left(m\right)}$,
of size $\left(2M+1\right)\times\left(2M+1\right)$, given by (\ref{eq:R_E}). 

Equation (\ref{eq:beam_forming_attenuation}) can be also written as in (\ref{eq:attenuation}), neglecting the noise terms.

\newpage

\begin{strip}
\begin{equation}
\mathbf{R}_{E}=\left[\begin{array}{lcr}
\left|\frac{E^{\left(-M\right)}}{E_{R}^{\left(-M\right)}}\right|^{2} & ... & \frac{E^{\left(-M\right)}}{E_{R}^{\left(-M\right)}}\,\left(\frac{E^{\left(M\right)}}{E_{R}^{\left(M\right)}}\right)^{*}\\
\frac{E^{\left(-M+1\right)}}{E_{R}^{\left(-M+1\right)}}\,\left(\frac{E^{\left(-M\right)}}{E_{R}^{\left(-M\right)}}\right)^{*} & ... & \frac{E^{\left(-M+1\right)}}{E_{R}^{\left(-M+1\right)}}\,\left(\frac{E^{\left(M\right)}}{E_{R}^{\left(M\right)}}\right)^{*}\\
... & ... & ...\\
\frac{E^{\left(M-1\right)}}{E_{R}^{\left(M-1\right)}}\,\left(\frac{E^{\left(-M\right)}}{E_{R}^{\left(-M\right)}}\right)^{*} & ... & \frac{E^{\left(M-1\right)}}{E_{R}^{\left(M-1\right)}}\,\left(\frac{E^{\left(M\right)}}{E_{R}^{\left(M\right)}}\right)^{*}\\
\frac{E^{\left(M\right)}}{E_{R}^{\left(M\right)}}\,\left(\frac{E^{\left(-M\right)}}{E_{R}^{\left(-M\right)}}\right)^{*} & ... & \left|\frac{E^{\left(M\right)}}{E_{R}^{\left(M\right)}}\right|^{2}
\end{array}\right]
\label{eq:R_E}
\end{equation}

\begin{equation}
y\left(t,\mathcal{S}\right)=\left\{ \begin{array}{ll}
E_{R}^{\left(0\right)}\,\sum_{m=-M}^{+M}w_{m}^{*}\,\frac{d_{0}}{d_{m}}\,\exp\left\{ -j\left(M-1\right)\frac{2\pi}{\lambda}d_{a}\frac{\cos\left\{ \left(\gamma+\varphi_{m}\right)/2\right\} }{\cos\left\{ \left(\gamma-\varphi_{m}\right)/2\right\} }\right\} +n_{m} & \textrm{if \ensuremath{\mathcal{S}=0}}\\
E_{R}^{\left(0\right)}\,\sum_{m=-M}^{+M}w_{m}^{*}\,\frac{E^{\left(m\right)}}{E_{R}^{\left(m\right)}}\,\frac{d_{0}}{d_{m}}\,\exp\left\{ -j\left(M-1\right)\frac{2\pi}{\lambda}d_{a}\frac{\cos\left\{ \left(\gamma+\varphi_{m}\right)/2\right\} }{\cos\left\{ \left(\gamma-\varphi_{m}\right)/2\right\} }\right\} +n_{m} & \textrm{if \ensuremath{\mathcal{S}=1}}
\end{array}\right.\label{eq:y_t}
\end{equation}

\begin{equation}
\begin{array}{l}
	A_{T}=\frac{\left|\sum_{m=-M}^{+M}w_{m}^{*}\,\frac{d_{0}}{d_{m}}\,\exp\left\{ -j\left(M-1\right)\frac{2\pi}{\lambda}d_{a}\frac{\cos\left\{ \left(\gamma+\varphi_{m}\right)/2\right\} }{\cos\left\{ \left(\gamma-\varphi_{m}\right)/2\right\} }\right\} \right|^{2}}{\left|\sum_{m=-M}^{+M}w_{m}^{*}\,\frac{d_{0}}{d_{m}}\,\exp\left\{ -j\left(M-1\right)\frac{2\pi}{\lambda}d_{a}\frac{\cos\left\{ \left(\gamma+\varphi_{m}\right)/2\right\} }{\cos\left\{ \left(\gamma-\varphi_{m}\right)/2\right\} }\right\} \,\left[1-j\frac{d_{m}}{\lambda}\intop_{S}\frac{1}{r_{1,m}r_{2,m}}\exp\left\{ -j\frac{2\pi}{\lambda}\bigl(r_{1,m}+r_{2,m}-d_{m}\bigr)\right\} d\xi_{2}d\xi_{3}\right]\right|^{2}}
\end{array}
\label{eq:attenuation}
\end{equation}
\end{strip}

\begin{center}
\begin{figure}
\begin{centering}
\includegraphics[scale=0.38]{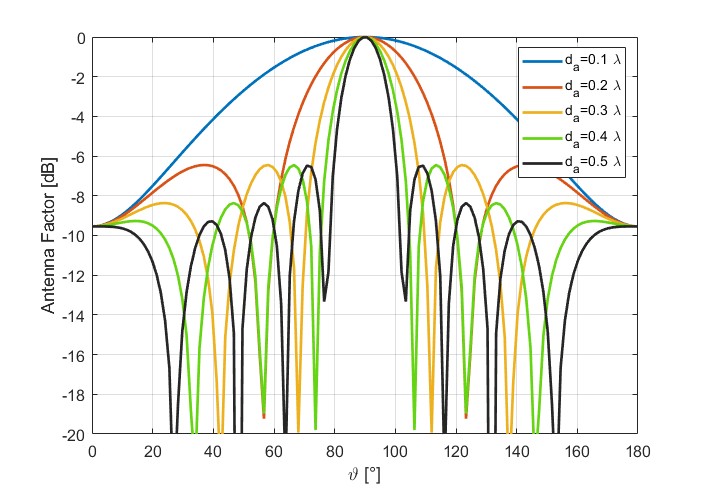} 
\par\end{centering}
\caption{\label{fig:array_factor}Array factor for an ULA composed by $9$ antennas ($M=4$) uniformly spaced from $0.1\,\lambda$ up to $0.5\,\lambda$.}
\end{figure}
\par\end{center}

\subsection{Array factor analysis}
The array factor $F_{a}=\mathbf{w}^{T}\mathbf{a}\left(\gamma\right)$
defines the response of the array as a function of the selected beamforming coefficients $\mathbf{w}$ and of the incident angle $\gamma$ of the steering vector $\mathbf{a}\left(\gamma\right)$ that characterizes the impinging wavefront. For instance, if (\ref{eq:steering_vector}) holds true with the weighting vector $\mathbf{w}$ such that, for all components $w_m$, it is $w_m=1/(2M+1)$, then the array factor is given by:

\begin{equation}
\begin{array}{ll}
F_{a} & =\mathbf{w}^{T}\mathbf{a}\left(\gamma\right)=\frac{1}{2M+1}\sum_{m=-M}^{+M}\exp\{{jm\frac{2\pi}{\lambda}d_a\cos\gamma}\}\\
 & =\frac{1}{2M+1}\cdot\frac{\sin\bigl\{\bigl(2M+1\bigr)\frac{\pi d_a}{\lambda}\cos\gamma\bigr\}}{\sin\bigl\{\frac{\pi d_a}{\lambda}\cos\gamma\bigr\}}.
\end{array}\label{eq:array_factor}
\end{equation}

Fig.~\ref{fig:array_factor} shows the modulus of the array factor in dB for an ULA of $9$ antennas ($M=4$) uniformly spaced with $d_{a}$ from $0.1\,\lambda$ up to $0.5\,\lambda$. It is apparent the effect due to the side lobes. The width of the first lobe $\Delta_{BW}$ is given by:

\begin{equation}
\Delta_{BW}=2\left|\arccos\left\{ \frac{\lambda}{\left(2M+1\right)d_{a}}\right\} \right|\approx\frac{2\lambda}{\left(2M+1\right)d_{a}},
\end{equation}

\noindent where the approximation holds for long arrays i.e. for $L=\left(2M+1\right)\,d_{a}\gg\lambda$. 

\begin{figure}[t]
\begin{centering}
\includegraphics[scale=0.60]{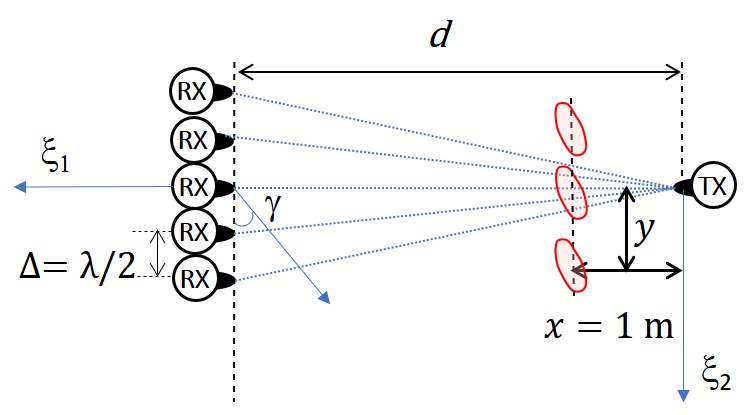}
\par\end{centering}
\caption{\label{fig:layout5}Link layout used for the simulations: the antenna array is composed by 5 antennas ($M=2$) while the target $S$, with size $a_z=0.9$ m and $a_y=0.275$ m, is placed in different positions along the line having distance $x=1$ m from the TX.}
\end{figure}

\begin{figure}[t]
\begin{centering}
\includegraphics[scale=0.31]{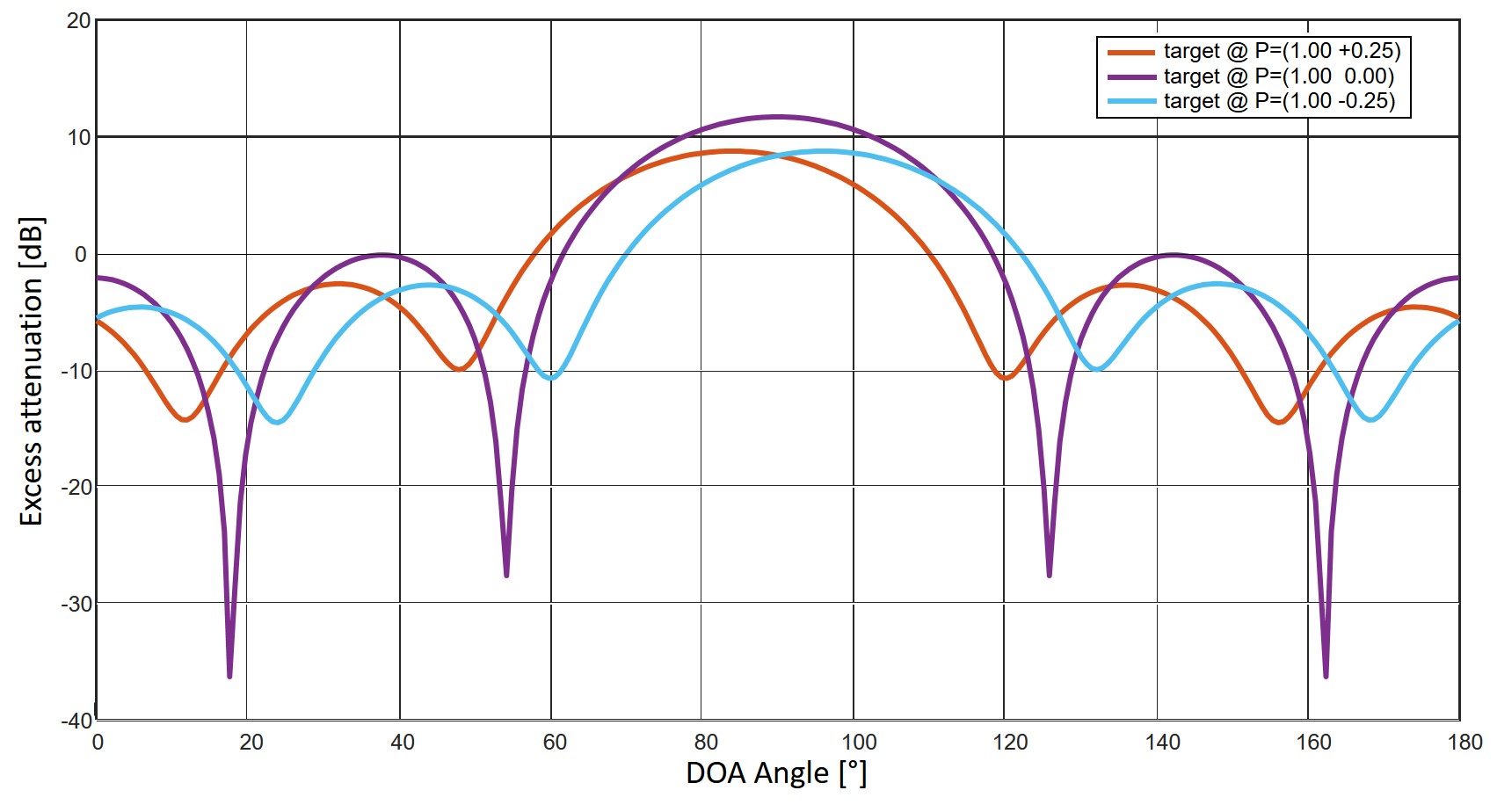}
\end{centering}
\caption{\label{fig:beams1}From top to bottom: a) target placement used for this test; b) excess attenuation expressed in terms of the DoA $\gamma$ of the array when the target $S$ is placed along the segment having distance $x=1$ m from the TX. Three different positions of the target $S$ are simulated at $(x,y)$: $(1.0,-0.25)$, $(1.0,0.0)$, and $(1.0,0.25)$; c) excess attenuation as observed on each single antenna without any array processing and for corresponding positions of the targets as above.}
\end{figure}

\section{Preliminary results}
\label{sec:preliminary_results}
In this section, we provide some preliminary results concerning the use of the array model of Sect.~\ref{sec:Array-based-body-model} vs the single-antenna model~\cite{rampa-2017}. 

The carrier frequency $f_c$ is set to $f_c=2.4868$ GHz, and we used an UL array of $5$ omni-directional antennas ($M=2$) spaced at $d_a=\lambda/2$ as shown in the layout sketch described in Fig.~\ref{fig:layout5}. The length of the central link of the array (i.e., for $m=0$) is equal to $d=d_0=4$ m while all links of the array are horizontally placed at height $h=0.9$ m from the ground.

As far as the body model is concerned, the absorbing 2-D sheet that represents the target has size $a_z=0.9$ m and $a_y=0.275$ m (i.e. a total size of 1.80 m  x 0.55 m). The target stands vertically on the floor, that is used only to support the target and does not have any EM influence on the radio links. The LoS of the central link will be the reference LoS line for the target positions with the origin of the axes placed in the TX as shown in Fig.~\ref{fig:layout5}.

\begin{figure}
\begin{centering}
\includegraphics[scale=0.31]{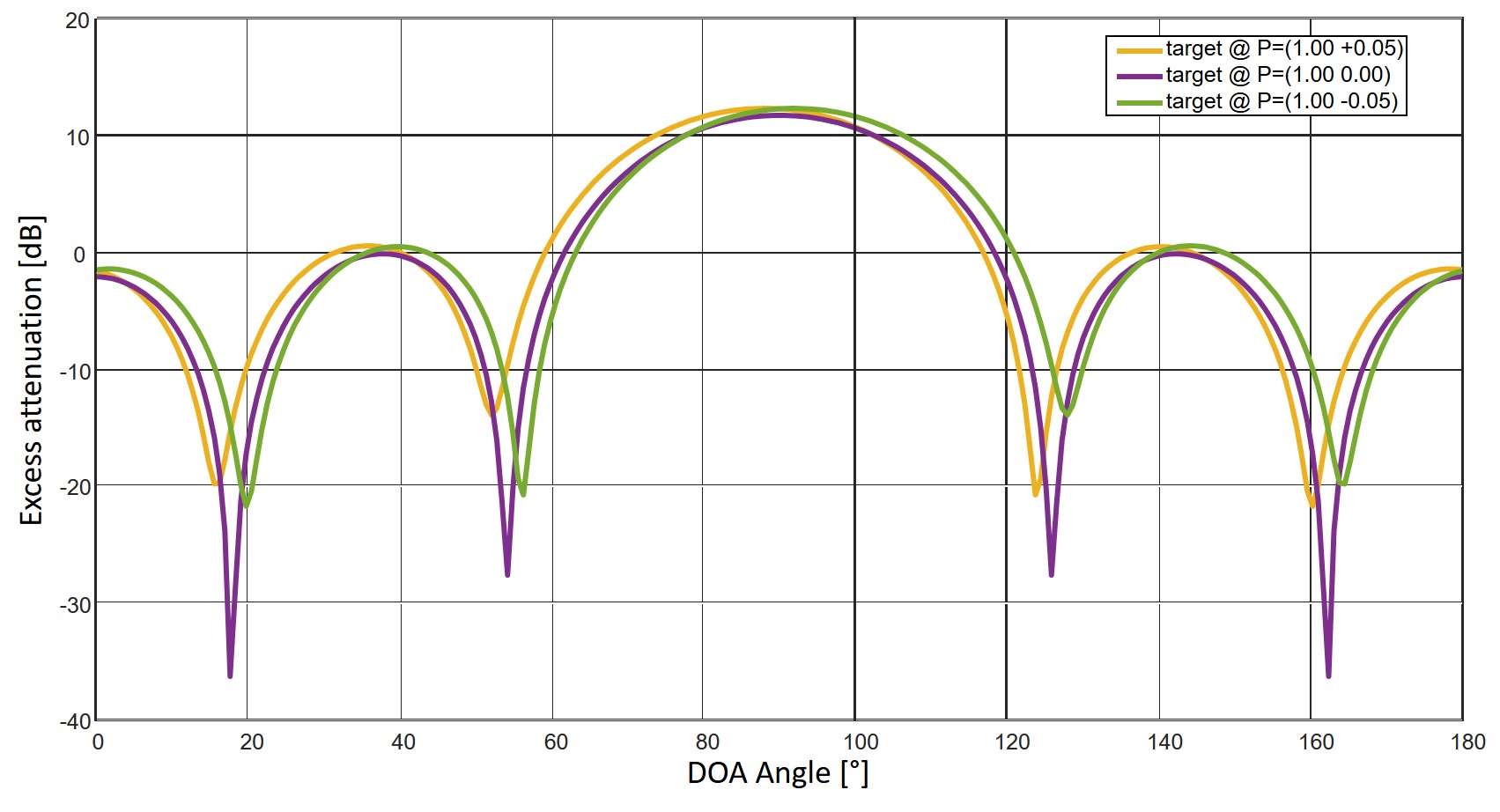}
\par\end{centering}
\caption{\label{fig:beams2}From top to bottom: a) target placement used for this test; b) excess attenuation of the array expressed in terms of the DoA $\gamma$ when the target $S$ is placed along the segment having distance $x=1$ m from the TX. Three very close positions of the target $S$ are simulated at $(x,y)$: $(1.0,-0.05)$, $(1.0,0.0)$, and $(1.0,0.05)$.}
\end{figure}

\begin{figure}
\begin{centering}
\includegraphics[scale=0.31]{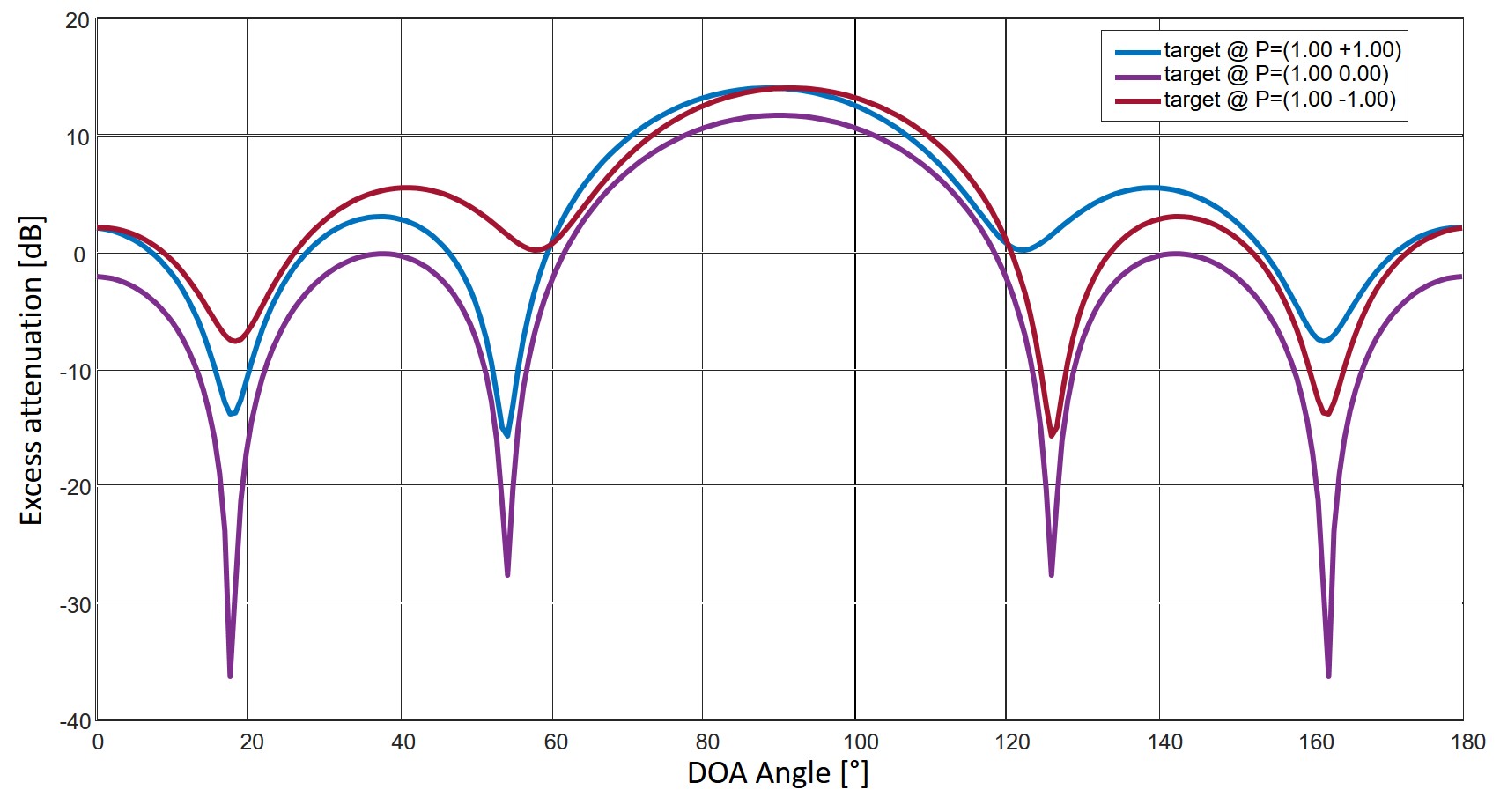}
\par\end{centering}
\caption{\label{fig:beams3}From top to bottom: a) target placement used for this test; b) excess attenuation of the array expressed in terms of the DoA $\gamma$ when the target $S$ is placed along the segment having distance $x=1$ m from the TX. Three widely spaced positions of the target $S$ are simulated at $(x,y)$: $(1.0,-1.0)$, $(1.0,0.0)$, and $(1.0,1.0)$.}
\end{figure}

Figs.~\ref{fig:beams1}, ~\ref{fig:beams2}, and ~\ref{fig:beams3} show the behavior of the entire array, namely the array response, in terms of the excess attenuation $20\,\log_{10}|E_R^{(0)}/E^{(m)}|$ as a function of the DoA $\gamma$ and for different values of the $y$ displacement  of the target $S$ (w.r.t. the central LoS). The locations of the target are limited to belong to the line at $\xi_1=x=1$ m that is orthogonal to the central LoS. Processing of the array signals to extract the response for varying DoA $\gamma$ is based on Fast Fourier Transform (FFT) and employs $N_{FFT}=257$  points.

For positions of the target $S$ near the LoS as in Figs.~\ref{fig:beams1}.a and~\ref{fig:beams2}.a, the extra attenuation for off-LoS positions (blue and yellow lines) are comparable with the one (red line) representing the target on the LoS (i.e., $\sim15$ dB for $y=0$ m compared with $\sim7$ dB for $y=\pm 0.25$ m). In addition, in Fig.~\ref{fig:beams1}.b the three positions, spaced by $0.25$ m, are clearly separable since they result in different array responses. Considering the closed-spaced positions (i.e., $5$ cm) of Fig.~\ref{fig:beams2}.b, the resulting extra attenuation values ($\sim17$ dB) are now much closer to the one of target placed on the LoS, thus making more difficult to visually separate these locations.
On the other hand, the targets are still distinguishable as they are associated with separable DoAs.

To compare the excess attenuation results for the antenna array of Fig.~\ref{fig:beams1}.b vs the ones of the single antennas, Fig.~\ref{fig:beams1}.c show the excess attenuation for each antenna. Comparing for instance the results for the target on the LoS (red line), it is apparent that, Fig.~\ref{fig:beams1}.c, the attenuation ranges from $\sim 14$ dB to $\sim 16$ dB depending on the selected antenna, while in Fig.~\ref{fig:beams1}.b the maximum value of the excess attenuation is the averaged value of $\sim 15$ dB.

For positions of the target $S$ outside the link area (i.e., outside the first Fresnel's ellipsoid, whose minor axis is $\sim 0.7$ m for the considered $f_c$), the influence due to the target is negligible. Some minor effects are visible in the nulls among the main and the side lobes, but they may be too weak to be exploited to identify the presence of the target in a real-world scenario, where there is always a certain degree of multipath.

We have also performed  some preliminary evaluations of conventional~\cite{benesty-2021} and MVDR (Minimum Variance Distortionless Response)~\cite{stoica-2003} beamforming algorithms. However, for space constraints, the results will not be discussed in this paper. Although preliminary, the results are promising and highlight the possibility of using EM body models to develop ad-hoc array processing tools adapted for passive DFL applications.

\section{Conclusions}
\label{sec:conclusions}
In this paper, we presented a tool for the simulation of the Electro-Magnetic (EM) effects of different body (i.e., target) positions inside a link area covered by an array of closely spaced omni-directional antennas. In the proposed preliminary study, we assessed the effects of the body on the antenna array response. In particular, we showed that for different target positions in the monitored area, the array is able to separate the contribution of the target when it is inside the first Fresnel's ellipsoid. The effects of the target can be represented both in terms of excess attenuation, which depends on body relative position, and as alterations of the array response pattern which loosely depends on target Direction of Arrival (DoA), relative to the Line-of-Sight (LoS) path. The proposed method paves the way for future applications of multiple antenna processing techniques in device-free localization applications.

\end{document}